\documentclass[aps,pra,twocolumn,groupedaddress,superscriptaddress,showpacs]{revtex4}

\usepackage{graphicx}
\usepackage{amsmath,amssymb,amsfonts}
\usepackage{bm}

\newcommand{\be}{\begin{equation}}
\newcommand{\ee}{\end{equation}}
\newcommand{\ba}{\begin{eqnarray}}
\newcommand{\ea}{\end{eqnarray}}

\begin{document}

\title{Reconciling quantum trajectories and stationary quantum
distributions\\ in single-photon polarization states}

\author{Alfredo Luis}
\affiliation{Departamento de \'{O}ptica, Facultad de Ciencias
F\'{\i}sicas, Universidad Complutense, 28040 Madrid, Spain}

\author{\'{A}ngel S. Sanz}
\affiliation{Instituto de F\'{\i}sica Fundamental (IFF--CSIC),
Serrano 123, 28006 Madrid, Spain}
\affiliation{Department of Physics and Astronomy, University College
London, Gower Street, London WC1E 6BT, United Kingdom}

\date{\today}

\begin{abstract}
The question of the representation of quantum stationary partially
polarized waves as random superpositions of different polarization
ellipses is addressed. To this end, the Bohmian formulation of
quantum mechanics is considered and extended to quantum optical
polarization. As is shown, this approach properly combines definite
time-evolving trajectories with rigorous stationary quantum
distributions via the topology displayed by the associated phase
field.
\end{abstract}


\maketitle


\section{Introduction}
\label{sec1}

In most courses, textbooks, and specialized treatises it is common
to introduce the polarization of harmonic waves as an ellipse
described by the electric field in a real configuration
representation \cite{tb}. Within this approach, partial polarization
thus arises by the rapid and random succession of more or less
different polarization states. In more advanced approaches, though,
the Stokes parameters and Poincar\'{e} sphere are introduced, which
allow us to assess the degree of polarization and make use of the
powerful tools provided by the SU(2) group. Moreover, the Stokes
parameters also enable a simple connection to the quantum regime in
terms of their quantum counterparts, namely, the Stokes operators
\cite{So}.

Once the Stokes parameters are introduced, usually the polarization
ellipses are left behind, so that the statistics of the electric
field in the polarization plane are abandoned. This is particularly
remarkable within the quantum domain, where uncertainty relations
imply that no field state can describe a perfect ellipse, just in
the same way that no particle can follow a definite trajectory
\cite{PMG,Q,Suso}. This thus suggests some kind of quantum
mismatching between the intuitive ellipse picture and the more
powerful Stokes-Poincar\'{e} formalism. More specifically, we find
quantum states with the maximum degree of polarization, but with
their electric-field distribution far from resembling any ellipse,
as seen, for example, in Fig.~\ref{fig2} below (for other examples,
see also Fig.~1 in Ref.~[\onlinecite{PMG}] or Fig.~4 in
Ref.~[\onlinecite{Q}]). Furthermore, the mismatch seems aggravated
for quantum stationary states (i.e., states with definite total
photon number). In such a case, the electric-field distribution is
constant even for states with the maximum degree of polarization,
which should correspond to an electric field rapidly describing an
elliptical trajectory.

In this work we investigate whether there is still a possibility to
keep the most intuitive approach, where partially polarized waves
are devised as random superpositions of different polarization
ellipses. To this end, the Bohmian formulation of quantum mechanics
\cite{BS} or, more properly speaking, its extension to optics
\cite{AP2010,JOSAA2012}, seems to be very convenient. This approach
properly combines the two ideas that we want to combine: definite
trajectories and rigorous quantum distributions. More specifically,
here we are going to study the Bohmian trajectories (or optical
paths) for a two-dimensional (2D) isotropic harmonic oscillator,
which properly represents in the quantum domain a transversal
two-mode harmonic wave. For definiteness, and for the sake of simple
illustration, we consider single-photon pure states. Accordingly, we
have organized this work as follows. In Sec.~\ref{sec2} we introduce
the basic elements involved in the representation of single-photon
polarization states. In Sec.~\ref{sec3} the associated Bohmian
dynamics is analyzed and discussed. Finally, in Sec.~\ref{sec4} the
main conclusions drawn from this work are summarized.


\section{Polarization of one-photon states}
\label{sec2}


\subsection{Polarization ellipse}
\label{sec21}

Consider a harmonic light wave consisting of two modes of the same
frequency, $\omega$, and with their corresponding electric fields
vibrating along orthogonal directions. These two modes are
represented by their complex amplitude operators $\hat{a}_1$ and
$\hat{a}_2$. This system is equivalent to a particle in a 2D
isotropic harmonic potential. The equivalence becomes clearer
through the quadrature operators representing the real and imaginary
parts of the electric field, $\hat{a}_j \propto \hat{X}_j + i
\hat{Y}_j$, with
\begin{equation}
 \hat{X}_j = \frac{1}{\sqrt{2}} \left ( \hat{a}^\dagger_j
  + \hat{a}_j  \right ) ,
 \quad
 \hat{Y}_j = \frac{i}{\sqrt{2}} \left ( \hat{a}^\dagger_j
  - \hat{a}_j  \right ) ,
\end{equation}
and commutator $[ \hat{X}_j , \hat{Y}_k ] = i\delta_{jk}$, with $j,k
= 1,2$. The $\hat{X}_j$ and $\hat{Y}_j$ operators are thus formally
equivalent, respectively, to the (dimensionless) position and
momentum of a 2D massive particle. For simplicity, and to exploit as
much as possible this equivalence, we assume $\hbar = 1$, $\omega =
1$, and $m=1$, with the latter being the mass of the equivalent
effective particle. The corresponding effective Hamiltonian then
reads as $\hat{\mathcal{H}} = \hat{a}^\dagger_1 \hat{a}_1 +
\hat{a}^\dagger_2 \hat{a}_2$.

The quadrature operators, $\hat{X}_j$ (with $j=1,2$), also allow us
to introduce a wave function for the 2D real electric field by
projection of the field state $| \psi \rangle$ on the unnormalized
joint eigenstate of the quadrature operators $\hat{X}_j
|x_1,x_2\rangle = x_j |x_1,x_2\rangle$. In particular, for
photon-number eigenstates, with $\hat{a}^\dagger_j \hat{a}_j
|n_1,n_2\rangle = n_j |n_1,n_2\rangle$, we have
\begin{equation}
 \langle x_1, x_2 | n_1, n_2 \rangle =
  \mathcal{N}_{12} H_{n_1} (x_1) H_{n_2} (x_2) e^{-(x_1^2 + x_2^2)/2} ,
\end{equation}
where $\mathcal{N}_{12} = (2^{n_1 + n_2}\pi n_1! n_2!)^{-1/2}$ is
the norm and $H_{n_j}$ are the corresponding Hermite polynomials.
Accordingly, the quantum analog of the polarization ellipse is the
distribution for the 2D real electric field \cite{PMG,Q,Suso},
\be
 P (x_1, x_2, t) = |\langle x_1 ,x_2 |\psi (t) \rangle|^2 ,
\ee
with $|\psi(t)\rangle = \exp (-i\hat{\mathcal{H}}t)
|\psi(0)\rangle$.


\subsection{Stokes parameters}
\label{sec22}

Regarding the Stokes picture of polarization, it can be well started
from the Stokes operators \cite{So},
\be
 \begin{array}{rcl}
  \hat{S}_0 & = & \hat{a}^\dagger_1 \hat{a}_1
   + \hat{a}^\dagger_2 \hat{a}_2 , \\
  \hat{S}_x & = & \hat{a}^\dagger_1 \hat{a}_1
   - \hat{a}^\dagger_2 \hat{a}_2 , \\
  \hat{S}_y & = & \hat{a}^\dagger_2 \hat{a}_1
   + \hat{a}_2 \hat{a}^\dagger_1 , \\
  \hat{S}_z & = & i ( \hat{a}_2 \hat{a}^\dagger_1 - \hat{a}^\dagger_2 \hat{a}_1) ,
 \end{array}
\ee
which satisfy the commutation relations
\begin{equation}
 \label{cr}
 [\hat{S}_x, \hat{S}_y ] = - 2 i \hat{S}_z ,
\end{equation}
with cyclic permutations, and
\begin{equation}
 \label{S2}
 \hat{\bf S}^2 = \hat{S}_0 (\hat{S}_0 + 2\hat{\mathbb{I}}),
  \qquad [ \hat{\bf S}, \hat{S}_0] = \hat{\bf 0} ,
\end{equation}
with $\hat{\bf S} = (\hat{S}_x, \hat{S}_y, \hat{S}_z )$. The
classical Stokes parameters are the mean values of the Stokes
operators, $s_j = \langle \hat{S}_j \rangle$. Because of the
non-vanishing commutator, as described by Eq.~(\ref{cr}), no state
can have definite values of all Stokes operators simultaneously (the
only exception being the two-mode vacuum, where they vanish
trivially). This is conveniently expressed by the uncertainty
relation
\ba
 (\Delta \hat{\bf S})^2 & = & (\Delta \hat{S}_x)^2
   + (\Delta \hat{S}_y)^2 + (\Delta \hat{S}_z)^2 \nonumber \\
 & = & \langle \hat{S}_0 (\hat{S}_0 + 2\hat{\mathbb{I}}) \rangle
   - \langle \hat{\bf S} \rangle ^2 \geq  2 \langle \hat{S}_0 \rangle ,
 \label{ur}
\ea
which holds after Eq.~(\ref{S2}) \cite{ur}, taking into account that
$|\langle\hat{\bf S} \rangle | \leq \langle \hat{S}_0 \rangle$.

\begin{figure}
\begin{center}
\includegraphics[width=8cm]{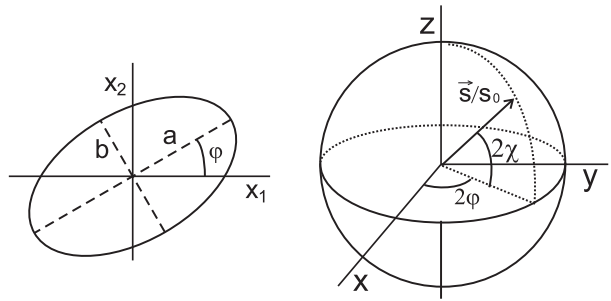}
\end{center}
\caption{\label{fig1} Polarization ellipse (left) and Poincar\'{e} sphere (right).}
\end{figure}

Consider now the standard (classical) definition of the degree of
polarization $\mathcal{P} = |\bm{s}| /s_0$. Some other more complete
definitions, though, have also been proposed, particularly within
the quantum domain \cite{Q,Suso,o}. Based on the fact that $|\bm{s}|
\leq s_0$, the Stokes parameters readily provide a representation
for polarization states in a unit sphere, namely, the Poincar\'{e}
sphere. This is done through the parametrization
\be
 \begin{array}{rcl}
  s_x & = & s_0 \cos ( 2 \chi) \cos ( 2 \varphi) , \\
  s_y & = & s_0 \cos ( 2 \chi) \sin ( 2 \varphi) , \\
  s_z & = & s_0 \sin ( 2 \chi) ,
 \end{array}
 \label{psp}
\ee
which is sketched in Fig.~\ref{fig1}. In the transformation
relations (\ref{psp}), $\tan ( 2 \varphi) = s_y /s_x$, with
$\varphi$ being the angle between the major axis of the polarization
ellipse and the $x_1$ axis. Regarding $\chi$, this angle determines
the ratio between the minor and major axes of the polarization
ellipse ($b$ and $a$, respectively), with $\pi/4 \geq \chi \geq -
\pi/4$ and $\tan \chi = \pm b/a$, where the sign is given by the
handedness (right-handed if $s_z >0$ and left-handed if $s_z <0$).
The relations (\ref{psp}) thus provide us with a one-to-one
correspondence between the Stokes parameters (or points on the
Poincar\'{e} sphere) and some average or mean polarization ellipse.

\subsection{One-photon states}
\label{sec23}

Let us now consider, more specifically, the most general pure
one-photon state. In the photon-number basis $|n_1,n_2\rangle$ it
reads as
\begin{equation}
\label{ops}
 |\psi\rangle = c_1 | 1,0 \rangle  + c_2  | 0, 1 \rangle ,
\end{equation}
with Stokes parameters
\be
 \begin{array}{rcl}
  s_0 & = & |c_1|^2 + |c_2|^2 = 1 , \\
  s_x & = & |c_1|^2 - |c_2|^2 , \\
  s_y & = & c_1^\ast c_2 + c_1 c_2^\ast
        = 2\ \! {\rm Re}(c_1 c_2^*) , \\
  s_z & = & i(c_1^\ast c_2 - c_1 c_2^\ast)
        = 2\ \! {\rm Im}(c_1 c_2^*) .
 \end{array}
\ee
These are all stationary states, i.e., $\hat{\mathcal{H}} |\psi
\rangle = | \psi \rangle$, so that $| \psi (t) \rangle = \exp (- i
t)| \psi (0) \rangle$. They are also SU(2) coherent states
\cite{cs}, which are usually considered as classical-like regarding
polarization \cite{sc}, as well as being minimum uncertainty states
of the uncertainty relation (\ref{ur}). Moreover, since $|\bm{s}|=
s_0$ for these states, we have $\mathcal{P}= 1$, thus displaying the
maximum degree of polarization according to the classic definition
seen above. In spite of this, they cannot be considered as having
perfect polarization, since $(\Delta \hat{\bf S} )^2 = 2 \langle
\hat{S}_0 \rangle \neq 0$. This is properly reflected by more
complete assessments of the degree of polarization \cite{Q,Suso,o}.

\begin{figure}
 \begin{center}
 \includegraphics[width=8.5cm]{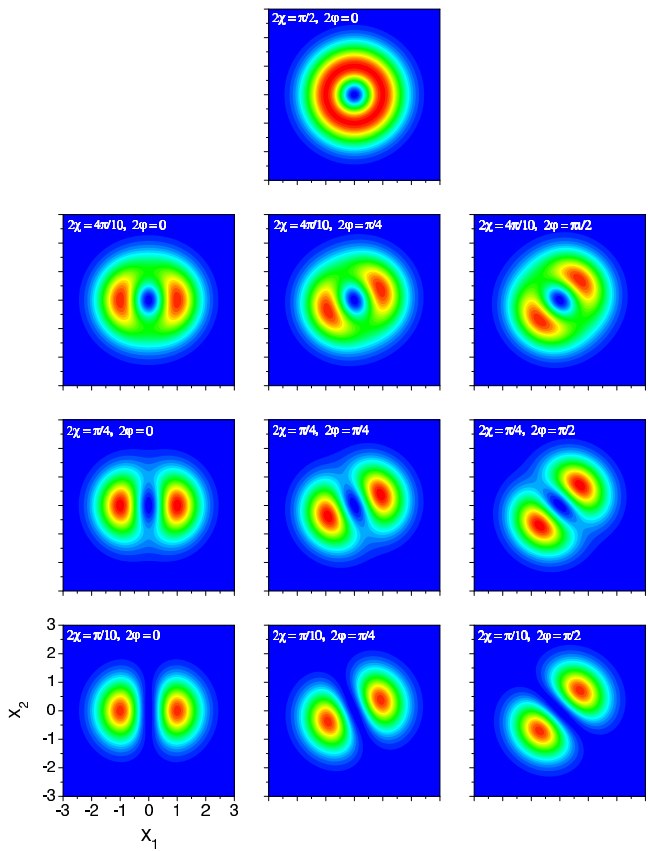}
 \end{center}
 \caption{\label{fig2} (Color online) Contour-plots corresponding
 to the $P(x_1,x_2)$ distribution of the 2D real electric field associated with the pure
  one-photon state (\ref{ops}).
  The values chosen for the angles $2\chi$ and $2\varphi$ cover the
  first quadrant of the Poincar\'e sphere (see Fig.~\ref{fig1}, right).
  Minimum to maximum is indicated with the transition from blue to
  red, starting from zero in all cases.}
\end{figure}

The quadrature wave function associated with the pure one-photon
state (\ref{ops}) is
\be
 \langle x_1, x_2 | \psi (t) \rangle =
 \sqrt{\frac{2}{\pi}}
   \left (c_1  x_1  + c_2 x_2 \right )
  e^{-(x_1^2 + x_2^2)/2 - it} ,
 \label{wf}
\ee
and the corresponding electric-field distribution is
\begin{equation}
 P (x_1, x_2 ) =
  \frac{1}{\pi} \left[ (1 + s_x) x^2_1 + (1 - s_x) x^2_2
   + 2 s_y x_1 x_2 \right] e^{-(x_1^2 + x_2^2 )} ,
 \label{pdef}
\end{equation}
which is effectively independent of time, i.e., it is {\it
stationary}, as expected. In Fig.~\ref{fig2} we show a series of
contour-plots of $P (x_1,x_2)$ for angles $2\chi$ and $2\varphi$
that cover the first quadrant of the Poincar\'e sphere; the behavior
in the remaining quadrants is analogous and can be inferred by
taking into account the corresponding symmetry relations. Also, for
convenience regarding the computation of the Bohmian trajectories
(see the discussion below), we have considered the minimum $2\chi$
to be $\pi/10$ instead of 0. As can be seen, only for $2\chi= \pm
\pi/2$ and any $\varphi$ does the picture of an ellipse remain valid
(this is for circularly polarized light). As the value of these
angles is changed, this picture breaks down despite having the
maximum degree of polarization in any of them ($\mathcal{P} =1$),
with the most dramatic breaking off being for $2\chi=0$, i.e., along
the equator of the sphere (this is for linearly polarized light). In
this case, $P(x_1,x_2)$ essentially consists of two separate lobes.
For $2\varphi=0$ these lobes are parallel to the $x_1$-axis (see the
lowest left panel), but their inclination changes as $2\varphi$
increases, thus becoming parallel to the $x_2$ axis when
$2\varphi=\pi$ (for the intermediate case, $2\varphi=\pi/2$, they
are $\pi/4$ inclined, as seen in the lowest right panel). Apart from
deviating from the ellipsoidal picture, it is also worth stressing
that information about the handedness will also get lost. This is
because $P(x_1,x_2)$ does not depend on $s_z$.


\section{Bohmian dynamics}
\label{sec3}


\subsection{Equations of motion}
\label{sec31}

Let us now consider the Bohmian approach with the purpose of
grasping some physical insight from the dynamical state associated
with the wave function (\ref{wf}). Within this context, the
corresponding Bohmian trajectories $\bm{r}(t)=[x_1(t),x_2(t)]$ are
obtained after integration of the guidance equation
\begin{equation}
\label{de}
 \dot{\bm{r}} = \nabla S = -i e^{-iS} \nabla e^{iS} ,
\end{equation}
where $S$ is the (real-valued) phase of the wave function
(\ref{wf}), i.e.,
\begin{equation}
 e^{i S} = \frac{(c_1 x_1 + c_2 x_2) e^{-it}}
  {\sqrt{|c_1|^2 x^2_1 + |c_2|^2 x^2_2
    + (c_1 c_2^* + c_1^* c_2) x_1 x_2}} .
\end{equation}
In terms of the Stokes parameters, this expression can also be
recast as
\be
 e^{iS} = \frac{\displaystyle \sqrt{1 + s_x}
 \left[x_1 + \left( \frac{1 - s_x}{s_y + is_z} \right) x_2\right]
  e^{-it + i\delta}}
 {\sqrt{x_1^2 + x_2^2 + s_x (x_1^2 - x_2^2) + 2 s_y x_1 x_2}} ,
\ee
where $\delta$ is a global relative phase associated with the
coefficient $c_1$. This phase factor is physically meaningless
regarding the topology displayed by the phase field, $S$, as well as
the trajectory dynamics described by the equation of motion below,
as also happens with $t$. Substituting this expression into
Eq.~(\ref{de}), we find
\be
 \label{gS}
 \dot{\bm{ r}} =
 \frac{s_z}{x_1^2 + x_2^2 + s_x \left( x_1^2 - x_2^2 \right)
  + 2 s_y x_1 x_2} \left(x_2 , -x_1 \right) .
\ee

In general, Eq.~(\ref{gS}) has to be numerically integrated in order
to obtain the corresponding trajectories, as seen in
Sec.~\ref{sec33} below. However, as shown in the next section, it is
also possible to draw a series of interesting conclusions directly
from the form of Eq.~(\ref{gS}), without the need to integrate it.


\subsection{Properties of the trajectories}
\label{sec32}


\subsubsection{Circular trajectories}

In spite of the complex dependence on $x_1$ and $x_2$ displayed by
the prefactor of the equation of motion (\ref{gS}), the trajectories
for any single-photon state with $s_z \neq 0$ are always circular.
This comes from the fact that
\begin{equation}
 \bm{r} \cdot \nabla S = \bm{r} \cdot \dot{\bm{r}} = 0 ,
 \label{circ}
\end{equation}
which implies that $\bm{r}^2$ (and therefore $\bm{r}$) is constant.

Now, for $s_z =0$, when the photons are linearly polarized, we have
$\dot{\bf r} \equiv {\bf 0}$ and therefore the whole vector $\bm{r}$
will be constant. That is, each separate ${\bf r}$-component is
constant with time, since Eq.~(\ref{circ}) ensures the time
independence only of $r = |{\bf r}|$, but not its components (see
below). This case can be then regarded as the limit of a circle
described at a vanishing speed.


\subsubsection{Nonuniform angular frequency}

The angular frequency of these circular trajectories is not uniform.
This is readily seen if we use polar coordinates,
\begin{equation}
 \label{pc}
 x_1 = r \cos \phi , \qquad  x_2 = r \sin \phi ,
\end{equation}
and then express (\ref{gS}) in terms of these coordinates, which
renders the following equation of motion for the polar component:
\ba
 \dot{\phi} & = &
  - \frac{s_z / r^2}{1 + s_x \cos 2 \phi + s_y \sin 2\phi}
  \nonumber \\
  & = & - \frac{\sin (2\chi) / r^2}
   {1 + \cos (2\chi) \cos (2\phi - 2\varphi)} .
 \label{velphi1}
\ea

We recall that the polarization ellipse is described at constant
angular speed $\omega$ (in our case $\omega =1$). However, as seen
through (\ref{velphi1}),
this is not the case for the Bohmian trajectories, where
$\dot{\phi}$ strongly depends on the polarization state (see below).
Thus, the only case with uniform angular velocity holds for $\chi =
\pm \pi/4$, i. e., for circularly polarized light.


\subsubsection{Handedness}

The handedness of the Bohmian trajectories is the same as for the
corresponding polarization ellipses. After the change of coordinates
(\ref{pc}), the Bohmian motion will be right-handed if $\dot{\phi} <
0$, while the polarization ellipse is right-handed when $s_z>0$.


\subsubsection{Extreme instantaneous frequencies\\ and ellipse
parameters}

The maximum instantaneous angular frequency of the Bohmian
trajectories, denoted by $\Omega = |\dot{\phi}|$, is
\begin{equation}
 \Omega_\mathrm{max} = \frac{1}{r^2 | \tan  \chi |} ,
\end{equation}
which holds when $\cos ( 2 \phi - 2 \varphi ) = -1$, i.e., when
$\phi = \varphi \pm \pi/2$, modulo $\pi$. In other words, this
condition is equivalent to saying that the maximum instantaneous
angular frequency of the Bohmian trajectory coincides with the
direction of the minor axis of the polarization ellipse (see
Fig.~\ref{fig1}).

On the contrary, the minimum instantaneous angular frequency is
\begin{equation}
 \Omega_\mathrm{min} = \frac{| \tan  \chi |}{r^2} ,
\end{equation}
which holds when $\cos ( 2 \phi - 2 \varphi ) = 1$, i.e., when
$\phi= \varphi$, modulo $\mathrm{\pi}$. That is,  the minimum
instantaneous angular frequency coincides with the direction of the
major axis of the polarization ellipse. Note, therefore, that the
ratio between minimum and maximum instantaneous angular frequencies
is
\begin{equation}
 \frac{\Omega_\mathrm{min}}{\Omega_\mathrm{max}}
  = \tan^2 \chi = \frac{b^2}{a^2} ,
\end{equation}
which coincides with the ratio of the minimum to maximum axes of the
polarization ellipse, $b$ and $a$, respectively.


\subsubsection{Compatibility with classical electrodynamics and non classicality}

We recall that the variables $x_1$ and $x_2$ are two orthogonal
electric-field components, so that the associated $x_1(t)$ and
$x_2(t)$ Bohmian trajectories represent the dynamical evolution of
the electric field. These orbits are {\it not} compatible with the
Maxwell equations, which demand that the electric field describes an
ellipse at constant angular frequency $\omega$ (in our case $\omega
=1$). The discrepancy between Bohmian orbits and classical
electromagnetism is apparent through Eq.~(\ref{gS}). This equation,
which expresses the time derivative of the electric field vector, is
strongly nonlinear. This is also clearly displayed in
Figs.~\ref{fig3}, \ref{fig4}, and \ref{fig5} below, where the
jumping behavior of the angular frequencies $\dot{\phi}$ are
irreconcilable with the harmonic behavior of classical polarization
ellipses.

We find this lack of compatibility with classical electrodynamics
very suggestive, since it may be naturally ascribed to the
nonclassical nature of stationary one-photon field states
(\ref{ops}). Of course, it can be readily shown that for non
stationary, classical-like Glauber coherent states the Bohmian
orbits are actually ellipses described at constant angular frequency
$\omega$, in full agreement with classical optics. However, the
mechanism that generates the dynamics in this case is very different
from the one involved in stationary single-photon states. In the
case of coherent states the dynamics appears because the phase of
the corresponding wave function is time dependent, while in the
latter it is just a purely topological property associated with
phase-local (time-independent) space variations. In this regard,
notice that the rich and nontrivial dynamics displayed by the
one-photon orbits is needed to reconcile electric-field dynamics
with the counterintuitive stationary electric-field probability
distribution associated with nonclassical photon-number light
states. This may provide a different perspective on the quantum
nature of light states.


\subsection{Numerical simulations}
\label{sec33}

\begin{figure}
 \begin{center}
  \includegraphics[width=8.5cm]{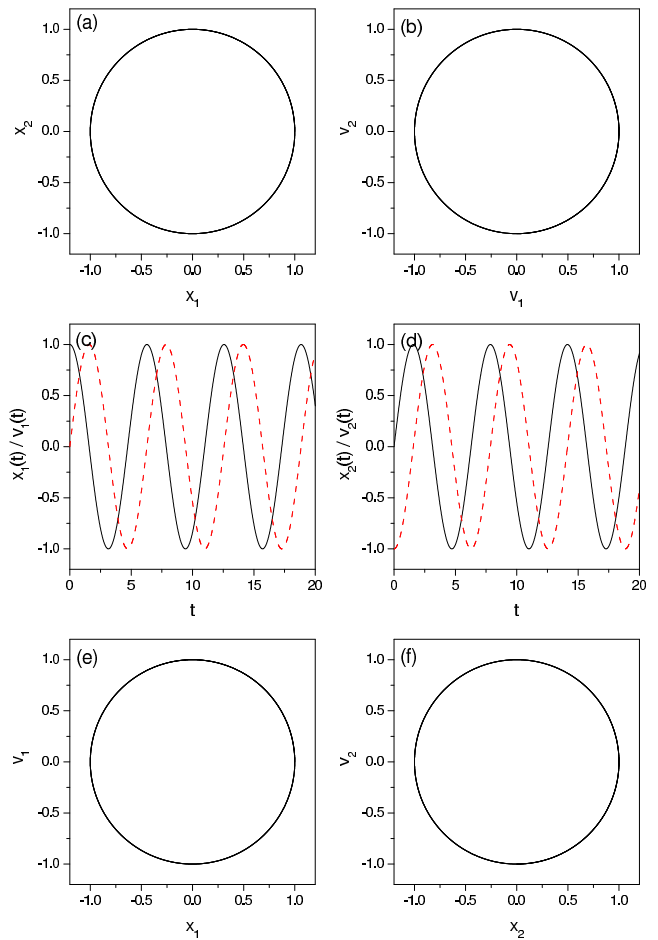}
 \end{center}
 \caption{\label{fig3} (Color online) Different representations
 of the Bohmian trajectories associated with a single-photon polarization state
  characterized by $2\chi=\pi/2$ and $2\varphi=0$ on the Poincar\'e
  sphere.
  In (c) and (d) the black solid lines denote the spatial
  component ($x_1$ or $x_2$), while the red dashed line refers to
  the respective velocity component ($v_1$ or $v_2$).
  In the corresponding panels, the velocity components have been
  obtained by evaluating (\ref{gS}) along the trajectory.}
\end{figure}

\begin{figure}
 \begin{center}
  \includegraphics[width=8.5cm]{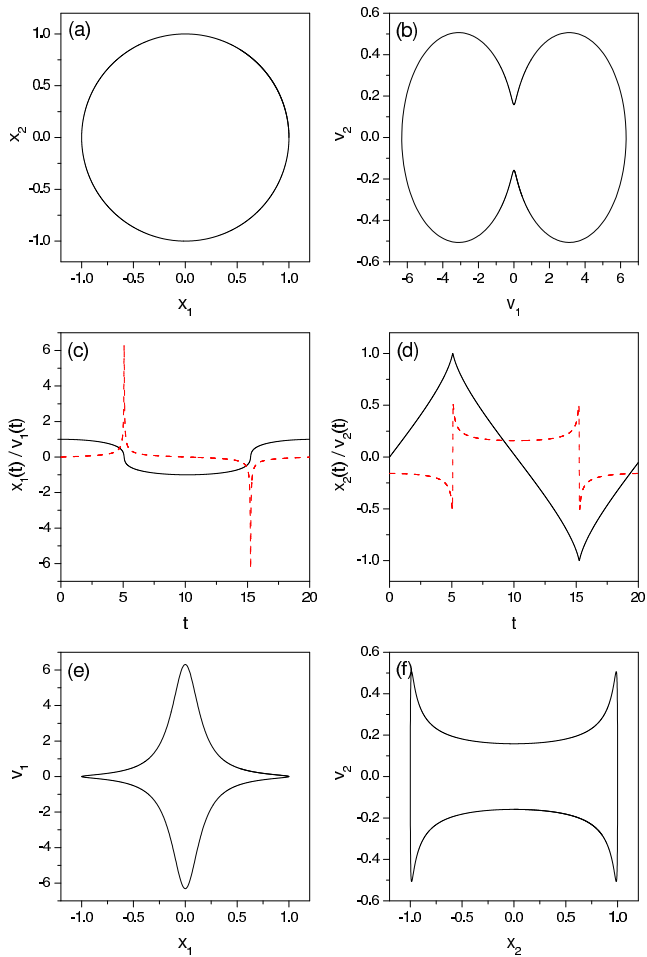}
 \end{center}
 \caption{\label{fig4} (Color online) As in Fig.~\ref{fig3}, but
 for $2\chi=\pi/10$ and $2\varphi=0$.}
\end{figure}

\begin{figure}
 \begin{center}
  \includegraphics[width=8.5cm]{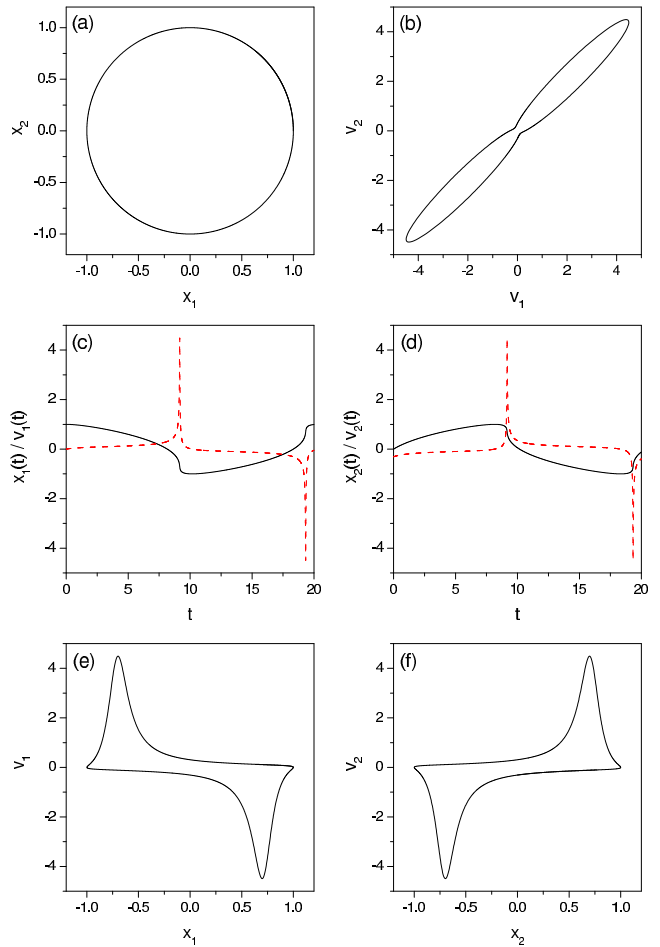}
 \end{center}
 \caption{\label{fig5} (Color online) As in Fig.~\ref{fig3}, but for
 $2\chi=\pi/10$ and $2\varphi=\pi/2$.}
\end{figure}

{\it A priori}, the properties discussed above may seem
counterintuitive and puzzling. For example, how is it possible that
the trajectories describe circles if their angular frequency is
variable with time, or if they tend to distribute in regions where
$P(x_1,x_2)$ is maximum, avoiding the nodal ones? Actually, it seems
that there is a mismatching between the symmetry of the wave
function and the features characterizing the trajectories. In order
to elucidate these questions as well as to reconcile two
antagonists, namely, motion and stationarity in the case of
single-photon states accounted for by wave functions like
(\ref{ops}), we have numerically integrated the equation of motion
(\ref{gS}) for some of the conditions considered in Fig.~\ref{fig2}.
Each condition represents a different single-photon state vector
and, therefore, a particular dynamics. The corresponding
trajectories as well as some other additional representations of
interest are displayed in Figs.~\ref{fig3}, \ref{fig4}, and
\ref{fig5}. For clarity, the initial condition, $x_1(0)=1$ and
$x_2(0)=0$, was chose to be the same in the three cases.
Accordingly, all the trajectories will be circles of unit radius
and, in principle, indistinguishable, in agreement with
(\ref{circ}). Disambiguation follows, though, when one analyzes the
time dependence of each component separately or their respective
velocities, as can be seen in the additional representations (see
also the discussions below). For completeness, and also to verify
our conclusions in other cases, we also considered the same cases,
but with other radii. The corresponding trajectories will not be
plotted here, although we would like to stress the fact that we
could corroborate that, effectively, the smaller the radius, the
faster the motion around the circle, in agreement with the $r^{-2}$
factor in Eq.~(\ref{velphi1}). Moreover, it has also been noticed
that the time needed for trajectories with the same radius to
complete the circle is different, increasing as $2\chi$ approaches
zero. This can be seen below by examining the dependence on time of
the velocity components of the trajectories, i.e., $v_i =
\dot{x}_i$, with $i=1,2$ [see the red dashed line in panels (c) and
(d) of Figs.~\ref{fig3}, \ref{fig4}, and \ref{fig5}]. Thus, as
$2\chi$ approaches zero, one observes a certain bistable behavior:
$v_1$ and $v_2$ undergo important variations in extremely short
periods of time, while they remain finite (or even vanishing) for
relatively long times, thus conferring a certain stationarity or
stability to the corresponding photon state. This fact is actually
in correspondence with the last of the properties studied in
Sec.~\ref{sec32}.

Going now to each particular case, in Fig.~\ref{fig3} we have
displayed the results for circularly polarized light $2\chi=\pi/2$
and $2\varphi=0$ on the Poincar\'e sphere (see Fig.~\ref{fig1}).
Notice that the evolution is ruled by a harmonic motion. This can be
seen through the time dependence of the components $x_1$ and $x_2$,
and their corresponding velocities $v_1$ and $v_2$ [see
Figs.~\ref{fig3}(c) and \ref{fig3}(d)], or through the respective
phase-space orbits [see Figs.~\ref{fig3}(e) and \ref{fig3}(f)]. In
this case, $s_z = s_0 = 1$ and $s_x = s_y = 0$, which means that the
photon has equal probability to be in the $|1,0\rangle$ state as in
the $|0,1\rangle$ state. Therefore, this uncertainty leads the
trajectory to visit all points on a circle of a given radius at the
same (angular) velocity, this eventually manifesting as a
harmonic-like motion or oscillation between $|1,0\rangle$ and
$|0,1\rangle$ (or, in other terms, between the horizontal and
vertical polarization states).

As we move towards the equator of the Poincar\'e sphere, i.e.,
vanishing $s_z$ for linearly polarized light (from top to bottom in
Fig.~\ref{fig2}), the torus-like distribution starts developing two
lobes, which in the limit $s_z=0$ become separate. Here we find an
apparently paradoxical behavior: $P(x_1,x_2)$ consists of two
separate lobes, but the trajectories are still circles, as indicated
by (\ref{circ}). To analyze this situation, we have proceeded as
before, but considering the single-photon state defined by
$2\chi=\pi/10$ and $2\varphi=0$ (see Fig.~\ref{fig4}). In order to
reconcile both behaviors, principally because no trajectory should
be expected at a nodal region (these are regions of lowest
probability), it is important to observe the time dependence of
$x_1$ and $x_2$, displayed in Figs.~\ref{fig4}(c) and \ref{fig4}(d),
respectively. In the case of $x_1$, we note a bistable behavior.
That is, $x_1$ is essentially $+1$ or $-1$ for relatively long
periods of time (which are expected to increase as $2\chi \to 0$),
while $x_2$ only oscillates linearly, up-and-down, between these
values to compensate the transition from $x_1$ to $x_2$. We observe
that such transitions are extremely fast. Actually, in the case of
$x_1$, the velocity $v_1$ is negligible except at the transition
times, while $v_2$ fluctuates between two almost stationary values
($+0.16$ and $-0.16$, approximately). As a consequence, the system
will remain apparently either in one lobe or the other of
$P(x_1,x_2)$, as shown by the orbit of the velocity representation
displayed in Fig.~\ref{fig4}(b) (if we consider $2\varphi \to \pi$,
we will find the same, but replacing $x_1$ by $x_2$, and vice
versa).

The situation is similar if we move around the equatorial plane
regarding the presence of the two lobes in $P(x_1,x_2)$, although
this motion implies a rotation of the axis along which they are
distributed. As mentioned in Sec.~\ref{sec23}, the lobes are aligned
with the $x_1$ axis for $2\varphi=0$ and start inclining
counterclockwise as this angle increases, becoming aligned with the
$x_2$ axis when $2\varphi=\pi$. This change affects not only the
distribution $P(x_1,x_2)$, but also the phase of the single-photon
state and therefore its associated trajectory dynamics. To
illustrate this case, we are going to consider the state described
by $2\chi=\pi/10$ and $2\varphi=\pi/2$ (see Fig.~\ref{fig5}), which
is linked to the distribution displayed in the lowest right panel of
Fig.~\ref{fig2}. Although the trajectory is still circular, the
orbit in the velocity space describes two lobes distributed along a
$\pi/4$ axis [see Fig.~\ref{fig5}(b)], which in the limit $2\chi \to
0$ will end up in a line with a $\pi/4$ inclination. Note that these
lobes constitute a distortion of those displayed by the velocities
in the previous case [compared with Fig.~\ref{fig4}(b)]. In
Figs.~\ref{fig5}(c) and \ref{fig5}(d) we observe that this behavior
just means that the photon oscillates between one lobe of the
distribution $P(x_1,x_2)$ and the other in a rather inhomogeneous
way: there is a slow, gradual approach to the transition followed by
a sudden switch, as indicated by the corresponding velocities. In
the corresponding phase-spaces [see Figs.~\ref{fig5}(e) and
\ref{fig5}(f)], we observe precisely that the velocity reaches its
maximum value right after the photon has reached one of the lobes
[of $P(x_1,x_2)$] and just before it jumps into the other.
Nonetheless, although the angular velocity is rather nonuniform, the
trajectory is still circular.

\begin{figure}
 \begin{center}
  \includegraphics[width=8.5cm]{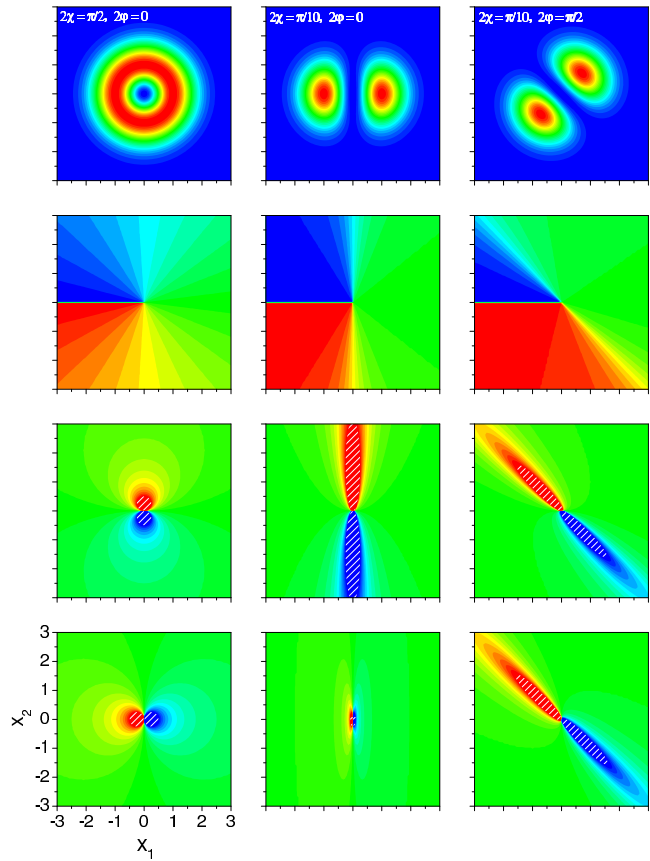}
 \end{center}
 \caption{\label{fig6} (Color online) Distribution $P(x_1,x_2)$ (top row),
  phase field $S$ (second row), and components of the local velocity field,
  $v_1$ and $v_2$ (third and fourth rows, respectively), for the
  polarization states considered in Figs.~\ref{fig3}, \ref{fig4},
  and \ref{fig5} (left, central, and right columns, respectively).
  Minimum to maximum is indicated with the transition from blue to
  red: for $P(x_1,x_2)$, starting from zero in all cases; for $S$,
  between $-\pi$ and $+\pi$; and for $v_1$ and $v_2$, between $-2$
  and $+2$.
  In the last two rows the shaded areas indicate a truncation to the
  maximum and minimum values chosen.}
\end{figure}

Finally, we would like to illustrate the source of motion in the
kind of stationary state considered here. In contrast to
non-stationary wave functions, where the time-dependence of the
phase leads naturally to time-evolving trajectories
\cite{CPL2007,JPA2008}, here motion has a topological origin: the
space-dependent (configuration space) gradient of the global phase
associated with the stationary state. In Fig.~\ref{fig6} we show the
distribution $P(x_1,x_2)$ (upper row), the phase $S$ (second row),
and the components of the local velocity field, $v_1$ and $v_2$ (in
the third and fourth rows, respectively), of the three cases
represented in Figs.~\ref{fig3}, \ref{fig4}, and \ref{fig5} (left,
central, and right columns, respectively). From a quick inspection,
the role of the phase-field topology on the trajectory dynamics
becomes readily apparent. This field decreases monotonically
counterclockwise, thus explaining the evolution of the trajectories
(from their beginning, as $x_1$ decreases, $x_2$ increases). Now,
while this decrease is gradual for $2\chi=\pi/2$, it becomes
step-like as $2\chi$ approaches zero. In the latter case, the step
lies precisely along the symmetry axis passing between the two
lobes. In this sense, the motion is relatively slow along each
plateau of the phase field and very fast when the trajectory comes
down the step (i.e., passes from the region covered by one of the
lobes to the region covered by the other). This has a
straightforward counterpart in the corresponding velocity field
components. For $2\chi=0$ these components are symmetric with
respect to a $\pi/2$ rotation, which is a signature of harmonic
motion (as one of the components starts to decrease, the other
increases, and vice versa). Now, as $2\chi$ approaches zero we find
two types of behaviors. For $2\chi=0$ and $2\varphi=0$, the
component along the step of the phase field becomes more prominent
than the component that is perpendicular to this step, which
eventually becomes meaningless. This is in correspondence with the
fact that it is more likely to find the trajectory either in the
right or the left lobe of $P(x_1,x_2)$ (small $v_1$ and $v_2$), but
not in between (large $v_1$). A similar behavior, but exchanging the
axes, would be found for $2\chi=0$ with $2\varphi= \pi$. Now, as
$2\varphi$ moves towards $\pi/2$, the two components tend to align,
thus becoming equal for $2\varphi=\pi/2$, which means that the
trajectories will avoid staying in the neighborhood of the diagonal
$x_2=-x_1$.


\section{Conclusions}
\label{sec4}

In this work we have shown that the topology displayed by the
Bohmian trajectories analyzed has no relation whatsoever with the
form of the polarization ellipse of the corresponding field state,
since they are always circles. However, the angular speed at which
these circles are described contains complete information about the
mean polarization state, i.e., the ellipse traditionally associated
with the Stokes parameters. This information is paradoxically absent
from the probability distribution corresponding to the 2D real
electric field [see Eq.~(\ref{pdef})], where one would expect to see
the quantum counterpart of the classical mean ellipse. Moreover,
seemingly, the Bohmian trajectories provide no hint about the
uncertainty of the polarization state, since the families of
trajectories associated with a given field state are essentially
identical.

Concerning the dynamics associated with single-photon polarization
states, we have found a relatively rich variety of behaviors
depending on the point on the Poincar\'e sphere defining the state
vector. As seen, although all the trajectories are circles, their
dynamical evolution is strongly dependent on their position with
respect to the Poincar\'e sphere, ranging thus from harmonic
behaviors (for trajectories on the poles) to bistable oscillating
ones (for trajectories on the equator). In all cases, though, this
dynamics has a clear topological origin associated with the
particular shape of the phase field, which goes from a gradual
decrease around $(x_1,x_2)=(0,0)$ (at the poles) to a step-like one
(on the equator). Accordingly, trajectories evolve counterclockwise
either harmonically or displaying jumps, respectively.

The fact that polarization stationary states may have an associated
internal or intrinsic dynamics is puzzling as well as challenging,
for it goes beyond our physical intuition even if it is proven
mathematically, as was done here. Even though distributions like
$P(x_1,x_2)$ are invariant objects with time, there is a dynamics
induced by the phase-field topology. This posses the interesting and
stimulating question of whether such an intrinsic motion could be
experimentally measured. Note that recently Steinberg and co-workers
were able to experimentally infer \cite{kocsis} the Bohmian
trajectories or averaged photon paths \cite{AP2010} in a two-slit
experiment by means of weak measurements \cite{aharonov,hiley}. The
ideas posed here clearly point towards a further extension of this
work in this direction with the ultimate goal of performing such
experimental measurements. This would allow us to better understand
the nature of stationary states as situations where the system is
statistically described by a time-independent (steady) distribution,
but that display particular inner dynamics. Obviously, this has a
potential interest in quantum information and quantum computation,
where the basic ingredient, the qubit, is precisely described by a
state like (\ref{ops}). Furthermore, also notice the implication in
quantum mechanics, where the role equivalent to polarization is
played by the particle spin. This vectorial quantity is
traditionally assigned to an internal rotation, because it fulfils
the same properties of the rotation group. With the picture on
stationary states provided above, based on a dynamical grounds, this
connection becomes even clearer.


\acknowledgments

Support from the Ministerio de Econom{\'\i}a y Competitividad
(Spain) under Projects No.~FIS2012-35583 (A.L.), No.~FIS2010-22082
(A.S.), and No.~FIS2011-29596-C02-01 (A.S.), and a ``Ram\'on y
Cajal'' Grant (A.S.); from the Consejer\'{\i}a de Educaci\'{o}n de
la Comunidad de Madrid under Project No.~QUITEMAD S2009-ESP-1594
(A.L.); and from the COST Action MP1006 ``Fundamental Problems in
Quantum Physics'' (A.S.) is acknowledged. A.S. also thanks the
University College London for its kind hospitality.


\end{document}